\documentstyle[12pt]{article}

\newcommand{\beq}{\begin{equation}}
\newcommand{\eeq}{\end{equation}}
\newcommand{\beqn}{\begin{eqnarray}}
\newcommand{\eeqn}{\end{eqnarray}}
\newcommand{\bi}{\bibitem}
\newcommand{\n}{\newline}

\newcommand{\la}{\langle}
\newcommand{\ra}{\rangle}
\newcommand{\bc}{\begin{center}}
\newcommand{\ec}{\end{center}}

\begin{document}

\pagestyle{empty}
\bc
{\Large {
{\bf Evidence of aging in spin glass mean-field models}
}}
\ec
\bc
{\large {
L. F. Cugliandolo, J. Kurchan
}}
\ec
\bc
{\large {
Dipartimento di Fisica, Universit\`a di Roma I,
{\it La Sapienza},
I-00185 Roma, Italy
}}
\ec
\bc
{\large {
INFN Sezione di Roma I, Roma, Italy
}}
\ec
\bc
{\large {
and
}}
\ec
\bc
{\large {
F. Ritort }} \footnote{Supported by Commission des
Communaut\'ees Europ\'eennes (Contract  B/SC1*/915198).
\n
Departament de Fisica Fonamental, Universitat de Barcelona, Diagonal 648,
08028 Barcelona, Spain.}
\ec
\bc
{\large {
Dipartimento di Fisica, Universit\`a di Roma II,
{\it Tor Vergata},
Via E. Carnevale, I-00173 Roma, Italy
}}
\ec
\bc
{\large {
October, 1993
}}
\ec
\vspace{0.05cm}

\bc
{\bf Abstract}
\ec

\vspace{0.3cm}

We study numerically the out of equilibrium dynamics
of the hypercubic cell spin glass in high dimensionalities.
We obtain evidence of aging effects qualitatively similar both to
experiments and to simulations of low dimensional models.
This suggests
that the Sherrington-Kirkpatrick model as well as
other mean-field finite connectivity lattices
can be used to study these effects analytically.

\vspace{0.5cm}
\noindent PACS numbers 75.10N, 75.40G, 75.50L

\newpage
\pagestyle{plain}

\setcounter{page}{1}

\baselineskip=25pt

There has been a long-standing and mainly unresolved controversy on which
kind of models can describe real (experimental) spin-glasses,
focused on whether the mean-field models (as opposed
to low-dimensional models) can or cannot describe, even at a qualitative
level, the essential features of them.
This controversy has been mainly centered
on the nature of the ground state structure
\cite{todos,Fisher}.

However, because of the long time scales involved in spin-glass dynamics,
most experiments are effectively performed out of equilibrium. Thus,
spin-glass physics is usually described as being essentially dynamical.
Indeed, the main point concerning the relevance of a given model should be
its dynamical behaviour - whether it resembles or not the experimental one.

Spin glasses exhibit the striking phenomenon of aging \cite{suecos}:
their dynamical properties
depend on their history even after very long times and they continue
to evolve long after thermalization at a
subcritical temperature.

A fully microscopic description of these effects in
realistic spin glasses is still lacking.
There have been several phenomenological attempts to describe the
physical mechanism of aging \cite{Fisher},\cite{PSAA}-\cite{Bouchaud}; and some
numerical studies of the three-dimensional Edwards - Anderson (3D EA)
model have
given results showing aging effects
in good agreement with experiments \cite{suecos2, Heiko1}.

Thus, the present understanding of the problem is at two extremes:
on the one hand, there are phenomenological models whose physics is
explicit but without a direct reference to the microscopy and,
on the other hand, there are simulations of microscopic models close
to real materials of which an analytical description seems at present
far off.

The interest of studying mean-field models is that they may provide a
bridge between the phenomenology and the microscopy of  real systems.
It was only recently shown that mean-field dynamical
models
can exhibit aging effects even in the thermodynamic limit \cite{CK}.
Although the model there considered (spherical spin glass with multi-spin
interactions) is simple enough that even analytic
results for the non-equilibrium dynamics
could be found, the price payed was that it is
quite unrealistic.

The scope of this article is to show
that the dynamical behaviour of the mean-field models
is strikingly similar to the behaviour of the
low-dimensional models and moreover that they mimic very well the
experimental observations.
With this aim we give evidence of
aging phenomena in a model whose behaviour is expected to
approach, for high dimensionality
$D$, that of high dimensional spin glasses on a hypercubic lattice and in
particular that of the Sherrington - Kirkpatrick (SK) model (as $D \rightarrow
\infty$).

The fact that mean-field models capture
the essential characteristics of spin glass experiments is promising because
they are technically
much simpler than realistic models.
Indeed, it has been argued that mean-field models can be solved
analytically whenever their long-term memory is weak; {\it i.e}
whenever the response to a constant field applied during a fixed
time-interval decreases to zero after long enough times \cite{CK}.
We will show below that this happens in this model.

\vspace{1cm}

In principle one would like to demonstrate aging effects in
the SK model, as the archetypical mean-field model.
However, the
use of Montecarlo dynamics is strongly limited
because of its full connectivity.
A whole sweep of the
lattice requires a computer time which grows as $N^2$
($N$ the number of spins) and this restricts the sizes that
can be analysed. As we shall discuss below, this implies
a strong limitation in the range of times
free of finite-size effects.
In order to study
mean-field dynamics up to large enough sizes
it is then convenient to choose lattices which are also
mean-field models
but with connectivity growing slower than $N$
(see, {\it e.g.} Ref. \cite{dom}).

The model we consider has been
introduced in Ref. \cite{felix}, where its equilibrium properties have been
studied.
It consists of a single hypercubic cell in $D$ dimensions;
on each of its corners there is a $\pm 1$ spin that interacts with its
$D$ nearest neighbours. The total number of spins is $N=2^D$. The
Hamiltonian is of the usual type
\[
H = - \sum_{\la ij \ra} J_{ij} \sigma_i \sigma_j
\label{hamiltonian}
\]
where $\la ij \ra$ denotes nearest neighbours
and the probability distribution of the couplings is given by
\[
P(J_{ij})
=
\frac{1}{2} \delta( J_{ij} - J)
+ \frac{1}{2} \delta( J_{ij} + J)
\; ,
\]
with $J=1/\sqrt D$.

This model is particularly interesting
since for large dimension $2D$ it is expected
to mimic the full lattice of dimension $D$ \cite{felix}
and to approach in the $D\rightarrow\infty$ limit the SK model
\cite{GMY}.
As regards to numerical simulations,
the computation time for a whole sweep of the cell
grows as $N \log N$.
Indeed,  Monte Carlo simulations of the SK
model for more than a few thousand spins become quickly unfeasible
while throughout this paper we consider $D=15,17$, {\it i.e}
$N= 2^{15}, 2^{17}$ spins.

Recently, Eissfeller and Opper \cite{opper} have devised a very
powerful Monte Carlo
procedure to solve the mean-field dynamical equations in the limit
$N \rightarrow \infty$, {\it i.e.} with no finite-size effects.
Unfortunately, this method requires a computer time that grows faster
than the square of the number of time-steps. As we discuss below, large
times are essential to distinguish aging phenomena
from an ordinary non-equilibrium relaxation.
Thus this semi-analytical
method would have required enormous computer
times to arrive at the times here considered.

We perform here a usual heat-bath dynamics. Even if the short-time
behaviour may  depend on the update procedure used, we expect long-time
features to be essentially independent of it. The computer time is linear
in the simulated time and, as we shall see, finite-size effects are not the
limiting factor.

We have simulated `field jump' experiments after a fast cooling to a
subcritical temperature which is afterwards kept constant and
measured the averaged magnetization:
\[
m(\tau)= \frac{1}{N} \sum_i \overline{ \la \sigma_i(\tau) \ra}
\]
and the correlation function
\begin{equation}
C(t_w+t,t_w)= \frac{1}{N} \sum_i \overline{ \la \sigma_i(t_w+t)
\sigma_i(t_w) \ra}
\; .
\label{corr}
\end{equation}
The overline denotes a mean over different
realizations of the couplings
and $\la \; \cdot \; \ra$ denotes an average over noise realizations.
The number of samples taken will be denoted  $N_s$. We will
present in detail the results for $T=0.2 \; T_c$,
($T_c = 1$),
and at the end we will indicate the dependence
of these results with the temperature.

In this
kind of simulations one should carefully select the appropriate time window
corresponding to the experimental physical situation.
For long enough times, a finite system eventually reaches equilibrium and
all aging effects dissapear. A necessary  (though not sufficient) condition
for having aging phenomena is that the time needed for the system to
achieve the thermodynamic (Gibbs-Boltzmann) distribution be longer than the
experimental time.
In a computer simulation of the kind we consider here $N$  is
finite, and one has to check that the observation times be small enough
that the system is not allowed to reach `thermal death' due to finite size.

The method we use to check this is to consider several copies of the
system with the same couplings starting from different random configurations
and evolving with different realizations of thermal noise.
We then calculate the evolution of the square of the
overlap between the configurations. This is a quantity that starts from
$O(1/N)$ and tends to $\overline{< q^2 >}_{eq}$, the mean square overlap
calculated with the equilibrium measure. Roughly,
$\overline{< q^2 >}_{eq} \simeq 0.7$  \cite{Giorgio}.
Since we only consider times such that the value of
$\overline{< q^2 >}(t)$ remains small
($\overline{< q^2 >}(t) \stackrel{<}{\sim} 0.04$),
the different copies are not able to cross barriers in their
search for the few deepest states.

A more subtle problem is that of small times: once one is satisfied that the
dynamics is non-equilibrium one has to check that  this is an {\em asymptotic}
aging process, {\it i.e.} not an ordinary out of equilibrium transient. In a
realistic system the question  is ultimately
resolved by the actual time-scales
involved, as compared with the experimental time. The model we are discussing
being only qualitatively realistic, we have to content ourselves with
some other criterion. We have chosen the following: we consider the asymptotic
non-equilibrium regime to start at the time when one-time quantities
(energy, magnetization, etc.) are near to their asymptotic values and
approach them with their asymptotic power law.
Because we are avoiding finite size effects,
`asymptotic' means large time limit taken {\em after} the large $N$ limit.

We have analysed the relaxation of the energy and performed a separate
power law fit for each time interval ($30-100$,
$100-300$, $300-1000$, $1000-3000$,
$3000-10000$ Monte Carlo sweeps (MCs)).
We found time exponents which were roughly consistent for
these intervals ($\simeq -0.3$) while the first $30$ steps
deviate from this behaviour.

We have simulated thermoremanent magnetization (TRM) and
zero field cooling (ZFC) experiments.
In the TRM experiments, the sample is rapidly
cooled from above the critical temperature
down to a temperature $T$ in the spin-glass
phase ($T < T_c$)
with a small field $h$ applied. Then, the system is allowed to evolve during
a `waiting' time $t_w$ at the constant temperature $T$ and field $h$.
After $t_w$ the
field is cut off and the relaxation of the magnetization $m(t+t_w)$, {\it i.e.}
the TRM, is measured as a function of the subsequent time $t$.
In the  ZFC experiments,
the sample is cooled from above $T_c$ in zero field and after a waiting time
$t_w$ a small field is applied.
Then, the increase of $m(t+t_w)$,
{\it i.e.} the ZFC magnetization, is measured.

The starting configuration in the simulations
was chosen at random corresponding to a fast quench to
a temperature $T<T^c$. This is slightly different from the experimental
procedures in that our scheme corresponds to a fast enough quench such that the
initial magnetization is still zero even in the TRM case. The difference is
however small, since the magnetization rapidly grows to its final value.
If linear response theory holds (as it should for small fields) one expects
that the sum of the magnetizations obtained from the TRM and the ZFC
processes with the same $t_w$ and $h$
yield the magnetization associated to a constant field $h$ applied
since the temperature quench (the field cooled magnetization (FCM)).
Because of the particular initial conditions, the sum of the ZFC plus
the TRM is not a constant, but a curve for the FCM that saturates very fast.
We have analysed the TRM, the ZFC
magnetization, their sum and the FCM for $t_w = 100$ and $t_w = 1000$,
$D=17$ and $h=0.1$: the linear response theory holds within $5\%$.
In Fig. 1 we show the plot for $t_w = 100$.
We henceforth concentrate only on TRM simulations.

In Fig. 2 we show the TRM simulations for $D=17$, $h=0.1$ and
waiting times $t_w = 100$, $300$, $1000$, $3000$, $10000$, with
$N_s = 10$ for $t_w = 100$, $300$ and $N_s = 5$ for the rest.
The curves clearly
depend on the waiting time: the response decreases with $t_w$. These
curves are very similar to, {\it e.g.},
the  corresponding experimental curves for the  indium-diluted chromium
thiospinel of Ref.  \cite{AHN}.

In Fig. 3 we show the decay of the correlation (eq.(\ref{corr}))
{\it vs.} $t$ for $D=15$, $h=0$, averaged over
$5$ samples. We note that the system distances from
itself with a speed that decreases with $t_w$: the phenomenon of `weak
ergodicity breaking' \cite{Bouchaud}. These
curves are remarkably similar to those of the 3D EA model \cite{Heiko1}.

In Fig. 4 we show the fitting for $t<<t_w$ of the correlation curves.
We plot $C(t+t_w,t_w) t^\alpha$ in terms
of $t/t_w$. The exponent that makes the curves superpose is
$\alpha=0.01$ and thus the
departure from a pure function of $t/t_w$ is very small.
Similarly there is a small
departure from a pure dependence on $t/t_w$ on the sector $t>>t_w$.

We have also performed simulations at higher temperatures (up to $T=0.8$).
As expected, aging effects are still present but they decrease
with increasing temperatures and they disappear as it
approaches the critical temperature.

\vspace{1cm}

In addition, we have analysed the response of the model
to changes in the temperature during the waiting time.

Firstly, we have performed
`temperature jump' simulations in the manner of
the experiments of Refs. \cite{suecos4}. The system has been
kept at a constant temperature $T-\delta T$ during $t_w$
when the temperature has been suddenly changed to $T$ and
afterwards it has been kept constant.
These results are presented in Figure 5.
We have measured the correlation function (\ref{corr}) for
various values of $\delta T$ and we have found the following:
\n
\indent If $\delta T > 0$ and $t>t_w$
the system behaves as a younger system and the greater the value
of $\delta T$, the younger the system seems to be. \n
\indent If $\delta T < 0$ and $t<t_w$ the system seems to be older and
the greater the absolute value of $\delta T$ the older the system
seems to be. \n
At times smaller than the waiting time ($t<t_w$) the response of the system
is the opposite. The $\delta T > 0$ ($\delta T < 0$) curve is below (above)
the reference $\delta T = 0$ curve. Furthermore, the responses are in this
range of times asymmetric as can be seen in the inset of Figure 5.
We do not have a full understanding of this point and to draw a
definite conclusion a more detailed analysis is needed.

Secondly, we have
performed some temperature cycling simulations in the manner of
the experiments in Refs. \cite{french,suecos3}.
We have found that the
system is very insensitive to short high temperature pulses during the
waiting time in the whole range of subcritical temperatures.
This last result seems to indicate that the total waiting times we have been
using are rather short: assuming the effect of a high temperature pulse is
somehow proportional to its absolute duration, then a small percentage of a
small waiting time necessarily has a small effect. If this is the case
one should not draw conclusions on the symmetry or asymmetry of the response to
small changes of temperature since one expects asymmetries to be pronounced
only at very long times.

\vspace{1cm}

The model studied has both frustration and disorder. It is interesting to
understand the relative importance of these two features as regards aging
effects.
With this aim we have studied the fully frustrated model \cite{derrida}
on the hypercubic cell. We have measured the correlation functions
(\ref{corr}) at constant temperature for various waiting times
and we have not observed any aging effect. However, we have found
that a very small amount of disorder introduced changing at random the
sign of a fixed small percentage of the bonds suffices to make the system
exhibit these effects.

\vspace{1cm}

In conclusion, there is at present good evidence of aging effects in
mean-field systems with ergodicity breaking in the thermodynamic limit:
`weak' and `true' ergodicity breaking coexist.
Our numerical results suggest that mean-field models have a
qualitatively similar dynamical behaviour to that of
low dimensional systems
({\it cf.} Refs. \cite{suecos2,Heiko1}).

\vspace{1cm}

We wish to thank C. De Dominicis,
J. Hammann, F. Lefloch, E. Marinari, M. Ocio,
G. Parisi, E. Vincent and  M. A. Virasoro for helpful
discussions.

\newpage

\baselineskip=25pt

{\bf Captions to Figures}
\vspace{0.5cm}

{\bf Fig. 1} Zero field cooled magnetization and thermoremanent magnetization
and their sum  for waiting time $t_w = 100$,
and the magnetization corresponding to a
constant field. For all the curves $D=17$, $T=0.2$ and $h=0.1$.
\vspace{0.5cm}

{\bf Fig. 2} Thermoremanent magnetization for waiting times
$t_w = 100$, $300$, $1000$, $3000$, $10000$,
$D=17$, $T=0.2$ and $h=0.1$.
\vspace{0.5cm}

{\bf Fig. 3} Autocorrelation function $C(t+t_w,t_w)$ {\it vs.} time $t$
for waiting times $t_w = 30$, $100$, $300$, $1000$, $3000$,
$D=15$ and $T=0.2$.
\vspace{0.5cm}

{\bf Fig. 4} Fitting (for $t<<t_w$) of the correlation functions
of Fig. 3. The plot shows
the autocorrelation functions times
$t^{-0.01}$  for each
waiting time {\it vs.} $t/t_w$.

{\bf Fig. 5} Autocorrelation function $C(t+t_w,t_w)$ {\it vs.} time $t$
in a `temperauture jump' simulation. $D=15$, $t_w = 3000$, $T=0.2$ and
$\delta T= \pm 0.01$.

\end{document}